\documentclass{article}
\usepackage{graphicx}
\usepackage{amsmath}
\usepackage{amsfonts}
\usepackage{amssymb}
\usepackage{amscd}

\def\beq{\begin{equation}}
\def\eeq{\end{equation}}

\begin{document}

\title{Spin and dynamics in relativistic quantum theories}

\author{W. N. Polyzou, \\
Department of Physics and Astronomy,
The University of Iowa, \\
Iowa City, IA 52242 \\
W. Gl\"ockle, \\
Institut f\"ur theoretische Physik II,
Ruhr-Universit\"at Bochum, \\
D-44780 Bochum, Germany\\
H. Wita{\l}a\\
M. Smoluchowski Institute of Physics, Jagiellonian
University, \\
PL-30059 Krak\'ow, Poland\\
}

\maketitle

\begin{abstract}
The role of relativity and dynamics 
in defining the spin and orbital angular momentum content of 
hadronic systems is discussed.  
\end{abstract}

\section{Introduction}
\label{intro}

There is a great deal of interest in the distribution of spin and
orbital angular momentum in hadronic systems.  In general the
underlying dynamics of partons in hadrons is relativistic.  In the 
relativistic case the coefficients that relate the parton spins to 
hadronic spins are momentum-dependent and are necessarily influenced
by the momentum-dependence of the hadronic wave function.  This 
momentum dependence appears in both relativistic quantum mechanics and 
relativistic quantum field theory.  In addition, in QCD, because 
constituent partons are confined, their ``mass'' becomes an additional 
dynamical variable. The purpose to this paper is to show how this 
dynamical dependence enters the relation between the hadronic and 
partonic spins and to show the equivalence of the treatment spin 
in Poincar\'e and Lorentz covariant theories.     

The treatment of spin in relativistic systems is different than it is in
non-relativistic systems.  In a relativistic system the spin of a
parton is identified with the angular momentum of the parton in its
rest frame while the spin of the hadron is defined as the angular
momentum of the hadron in its rest frame.  Transforming a parton from
its rest frame to the hadrons rest frame, where the spins can be
coupled, involves boosts which generate dynamical rotations.  These
rotations transform the parton spins and also impact the relative
orbital angular momentum before they can be coupled.  The spin of the
constituents, the internal orbital angular momentum and the spin of
the system are related by Clebsch-Gordan coefficients of the
Poincar\'e group\cite{moussa}\cite{bkwp}\cite{wgwp} .  The
Poincar\'e group Clebsch-Gordan coefficients are labeled by
eigenvalues of mass and spin Casimir operators, which are dynamical
operators.

An additional complication is that boosts to the rest frame are not
unique; a boost to the rest frame followed by a momentum-dependent
rotation is a different boost to the rest frame.  There are as many
different kinds of boosts as there are momentum-dependent rotations.
Each boost defines a different spin observable.  For example, there
are distinct boosts that are used to define the helicity, light front
spin or canonical spin.  These are three specific choices, that are
distinguished by useful properties, out of an infinite number of
possibilities.  In many-body systems there is another relevant spin,
which we call the constituent spin\cite{wgwp}, which is distinguished by the
property that spins and orbital angular momenta can be combined using
ordinary SU(2) coupling methods to get the hadronic spin.  

While all of the spins satisfy $SU(2)$ commutation relations, the
different spins observables are related by the momentum dependent
(Melosh) rotations\cite{melosh} that relate different boosts.  Because different
spin observables differ by momentum-dependent rotations, partial
derivatives with respect to momentum that hold one spin observable
constant will not commute with a different spin observable.  This
means not only are there an infinite number of possible spin
observables, but each one is associated with a different quantity that
can be identified with an orbital angular momentum.  As a result the
spin and orbital angular momentum content of a hadron is dynamical and
representation dependent.  In what follows we discuss some of the
relevant issues.

The dependence of the spin on the choice of boost is seen in the 
relations between the spin and angular momentum
\begin{equation}
j_a^l := {1 \over 2} \epsilon^{lmn} 
{B_a^{-1} (p)^m{}_{\mu}}
{B_a^{-1} (p)^n{}_{\nu}}J^{\mu \nu}
\label{eq:a1}
\end{equation}
where $B_a^{-1} (p)^m{}_{\mu}$ is a boost that maps $p$ to its rest frame.
Here $a$ is an index used to distinguish different types of Lorentz boosts.

This definition can be equivalently expressed in terms of the polarization 
vectors,  
$e_a^{m}{}_{\mu}(p):= B_a^{-1} (p)^m{}_{\mu}$, $m=1,2,3$, 
that are three orthonormal space-like vectors that are
orthogonal to the 
four momentum:
\begin{equation}
j_a^l = {1 \over 2} \epsilon^{lmn} 
{e_a^{m}{}_{\mu}(p)
e_a^{n}{}_{\nu}(p)} J^{\mu \nu} .
\label{eq:a2}
\end{equation}
Spins constructed using different boosts (labeled by $a$ and $b$) 
are related by momentum-dependent 
Melosh rotations  
\begin{equation}
j_a^l = R_{ab}(p)^l{}_m j_b^m \qquad
\mbox{where} \qquad 
R_{ab}(p)^l{}_m=
B_a^{-1} (p)^l{}_{\mu}
B_b  (p)^{\mu}_{m} .
\label{eq:a3}
\end{equation}
Because the different types of spin observables differ by momentum-dependent
rotations, 
``Position operators''  \cite{newton}\cite{wp}\cite{bkwp}that involve partial derivatives with respect to 
momentum need to specify which kind of spins is being held 
constant during the differentiation,  
\begin{equation}
[\pmb{\nabla}_{P_{\vert_{j_a}}}, \mathbf{j}_a]=0 \,
{\Rightarrow} \, [\pmb{\nabla}_{P_{\vert_{j_a}}}, \mathbf{j}_b] \not= 0 . 
\label{eq:a4}
\end{equation}
These partial derivatives can be written in terms of the Poincar\'e
generators with $\mathbf{V}= \mathbf{P}/M$ by \cite{wp}\cite{bkwp}
\begin{equation}
X^k_a= 
{i \pmb{\nabla}_{\mathbf{P}_{\vert_{\mathbf{j}_a}}}}
=
-{1 \over 2} \{H^{-1},{K}^k\} + i H^{-1} C_{1a}^{kl}(\mathbf{V}) j_a^l  
\label{eq:xx}
\end{equation}
where $H$ is the Hamiltonian, $M= \sqrt{H^2-\mathbf{P}^2}$ is the 
invariant mass operator and $\mathbf{V}$ is the four velocity.
In terms of these operators the spins and angular momentum are related by
\begin{equation}
J^j = (\mathbf{X}_a \times \mathbf{P})^j + C_{2a}^{jk} (\mathbf{V})j_a^k 
\label{eq:a5}
\end{equation}
where the operators $C_{1a}^{jk}(\mathbf{V})$ and 
$ C_{2a}^{jk} (\mathbf{\mathbf{V}})$ are the following functions 
of the Poincar\'e generators:
\begin{equation}
C_{1a}^{jk}(\mathbf{V}) =
{1 \over 2} \mathbf{Tr} [B_a(\mathbf{V})^{-1}\sigma_j B_a(\mathbf{V}) \sigma_k]
- V^0 \mathbf{Tr} [B_a(\mathbf{V})^{-1}{\partial \over \partial V_l}  
B_a(\mathbf{V}) \sigma_m]
\label{eq:a6}
\end{equation}
\begin{equation}
C_{2a}^{jk} ({\mathbf{V}}) = 
{1 \over 2} \mathbf{Tr} [B_a(\mathbf{V})^{-1}\sigma_j B_a(\mathbf{V}) \sigma_k]
+
i \epsilon_{jlm} 
\mathbf{Tr} [B_a(\mathbf{V})^{-1}{\partial \over \partial V_l}  
B_a(\mathbf{V}) \sigma_m] 
\label{eq:a7}
\end{equation}
and $B_a(\mathbf{V})$ is the $SL(2,\mathbb{C})$ representation of the 
$a$-boost.  The quantity $\mathbf{X}_a \times \mathbf{P}$ is the
associated orbital angular momentum \cite{wp}\cite{bkwp} 

Three components of the four momentum and the projection of any of these
spin observables on a given axis are labels for vectors in irreducible 
subspaces.  Products of two such irreducible representations can be 
expressed as direct integrals of composite irreducible representations
using the Clebsch-Gordan coefficients for the Poincar\'e group.
Like any set of Clebsch-Gordan coefficients, the actual coefficients
depend on the choice of irreducible basis.
The Poincar\'e group Clebsch-Gordan coefficients for a basis 
labeled by the $a$-type spin are 
\[
_a\langle  (M_1,j_1) \mathbf{P}_1 ,\mu_1  (M_2,j_2) \mathbf{P}_2 ,\mu_2 
\vert k, j (M_1,j_1,M_2,j_2) \mathbf{P} ,  \mu , l , s_{12} \rangle_a =
\]
\[
\sum_{\mu_1',\mu_2',\mu_1'',\mu_2'',\mu_s,m}
\delta (\mathbf{P} - \mathbf{P}_1-\mathbf{P}_2)
{\delta (k -k(\mathbf{P}_1,\mathbf{P}_2)) \over k^2}
\times
\]
\[
\sqrt{{\omega_{M_1} (\mathbf{k})\omega_{M_2} (\mathbf{k})
\over \omega_{M_1} (\mathbf{P}_1)\omega_{M_2} (\mathbf{P}_2)}}
\sqrt{{
\omega_{M_1} (\mathbf{P}_1)+\omega_{M_2} (\mathbf{P}_1)
\over \omega_{M_1} (\mathbf{k})+\omega_{M_2} (\mathbf{k})}}
\times
\]
\[
{D^{j_1}_{\mu_1 \mu_1'}[R_{wa} (B_a(V) ,k_1)]
D^{j_1}_{\mu_1'\mu_1''} [R_{ac}(k_1)]} \times
\]
\[
{D^{j_2}_{\mu_2 \mu_2'}[R_{wa} (B_a(V) ,k_2)]
D^{j_2}_{\mu_2'\mu_2''} [R_{ac}(k_2)]}
\times 
\]
\begin{equation}
Y^{l}_{m} (\hat{\mathbf{k}}(\mathbf{P}_1,\mathbf{P}_2)) 
\langle j_1, \mu_1'', j_2, \mu_2'' \vert s_{12}, \mu_s \rangle
\langle l, m, s_{12}, \mu_s \vert j, \mu \rangle
\label{eq:a8}
\end{equation}
These involve two types of spin rotations.  There are Wigner rotations
$R_{wa} (B_a(V) ,k_i)$ that arise from the $a$-boosts that relate the
system and parton rest frames and generalized Melosh rotations,
$R_{ac}(k_2)$, that transform the resulting spins to the canonical
spin representation where all of the spins and orbital angular momenta
Wigner rotate together so they can be added using ordinary $SU(2)$
spin addition.

The spins obtained by applying these two rotations to the hadronic spins
are the constituent spins mentioned earlier.  These are 
the spins associated with the magnetic quantum numbers
$\mu_i''$ in (\ref{eq:a8}).  It is apparent from
this equation that when these spins are combined with the orbital 
angular momentum using spherical harmonics and SU(2)
Clebsch-Gordan coefficients the result is the total spin.

The Poincar\'e group Clebsch-Gordan coefficients (\ref{eq:a8}) 
simplify in special bases.  If the
spins are defined using the standard rotationless boosts there are no
Melosh rotations, if the rotationless boost is replaced by a
light-front boost there are no Wigner rotations, and if the
rotationless boost is replaced by a helicity boost the Wigner
rotations become multiplication by a phase.

One result of the momentum-dependence of the rotations is that the
momentum-dependence of the hadronic wave function affects the
expectation values of both the spins and orbital angular momentum.
Since the momentum dependence of the hadronic wave function is
determined by the dynamics, the dynamics enters in the spin coupling
when the Poincar\'e Clebsch-Gordan coefficients are integrated against
the hadronic wave functions.

When one couples two interacting subsystems, one has to ask whether
the masses in the Poincar\'e Clebsch-Gordan coefficients are the
physical masses of the subsystems or the invariant masses of their
constituents.  For example, the mass of a meson or the invariant mass
of a quark antiquark pair?  So far we have treated them as invariant
masses of the constituents.  Cluster properties suggest that one
should really use the physical mass operators of the subsystems.
Fortunately there is a unitary transformation that removes the
interaction dependence from the hadronic spin\cite{fcwp}\cite{wp2}.
In this representation the spins can be coupled by sequential coupling
using the Clebsch-Gordan coefficients of the Poincar\'e group as if
the particles were not interacting.  This unitary transformation
changes the Hamiltonian, generating many-body interactions.  It also
changes the representation of the wave function in a way that
preserves probabilities, expectation values, as well as scattering
observables.  As a result of this all of the dynamical spin effects can be
absorbed by changing the representation of the wave function.
In QCD the quark masses themselves are also not constant.

A second ambiguity with spin has to do with whether the dynamics is
formulated using Poincar\'e covariant or Lorentz covariant bases.
Field theories are normally formulated using Lorentz covaraint bases
while relativistic quantum mechanics is typically formulated using
Poincar\'e covariant bases.  These are simply related; the dynamics
enters both representations, but in different but equivalent ways.  To
understand this note that the unitary representation of the Poincar\'e
group on positive-mass positive-energy irreducible basis states has
the form
\begin{equation}
U(\Lambda ,0) \vert (M,j)\mathbf{P},\mu \rangle_a = 
\sum \vert (M,j)\pmb{\Lambda}P,\nu \rangle_a 
D^j_{\nu \mu}(R_{wa}(\Lambda,P)).
\label{eq:a9}
\end{equation}
The Wigner rotation can be decomposed into the composition of a
boost followed by a Lorentz transformation followed by an 
inverse boost with the transformed four momentum
\begin{equation}
R_{wa}(\Lambda,P) = B_a^{-1}(\Lambda P) \Lambda B_a (P).
\label{eq:a10}
\end{equation}
The group representation property can be used to split the Wigner function
apart.  The finite dimensional representations of $SU(2)$ are related
to finite dimensional representation of $SL(2,\mathbb{C})$ by analytic 
continuation\cite{weinberg}\cite{wgwp}, so we can 
still use the group representation property.
Absorbing the Wigner functions of the boosts into the
states gives the Lorentz spinor representation of the states:
\begin{equation}
\vert (m,j) \mathbf{P}, {b} \rangle := \sum_{\mu}
\vert (m,j) \mathbf{P}, \mu \rangle_a 
D^j_{\mu {b}}[B_a^{-1}(P/M)].
\label{eq:a11}
\end{equation}
Here the boosts are represented by $2 \times 2 \, SL(2,\mathbb{C})$
transformations.  These spinor basis states (\ref{eq:a11}) 
have the following 
Lorentz covariant transformation property
\begin{equation}
{U(\Lambda, 0)} \vert (m,j) \mathbf{P}, {b} \rangle =
\sum_{{b}'} \vert (m,j) \pmb{\Lambda}{P}, {b}' \rangle
{D^j_{{b}'{b}}[\Lambda]}.
\label{eq:a12}
\end{equation}
The price paid for using the covariant representation is that
the Hilbert space inner product becomes dynamical
\[
\langle \psi \vert \phi \rangle =
\]
\begin{equation}
\int 
\langle \psi \vert (m,j) \mathbf{P}, {b} \rangle 
d^4P \theta(P^0) \delta (P^2 +
{M^2}) 
D^j_{bb'}[P^{\mu}\sigma_{\mu}/{M}]  
\langle (m,j) \mathbf{P}, {b}' \vert \phi \rangle 
\label{eq:a13}
\end{equation}
where we have used 
the hermiticity of the $SL(2,\mathbb{C})$ representation of the
rotationless boost which gives
\begin{equation}
B_a (V)B_a^{\dagger}(V) =
B_c (V)R_{ca}(V) R^{\dagger}_{ca} B_c^{\dagger}(V) = B_c (V)B_c^{\dagger}(V) = B^2_c (V)
= P^{\mu}\sigma_{\mu}/M
\label{eq:a14}
\end{equation}
independent of the type $(a)$ of boost.  The zero component of
$\sigma^{\mu}$ in (\ref{eq:a14}) is the identity and the other three
components are the Pauli matrices.  In (\ref{eq:a13}) the dynamics is
appears in the mass-shell condition, which makes the Wigner function
into a positive matrix.  The inner product (\ref{eq:a13}) is identical
to the original Poincar\'e covariant inner product.

Unlike representations of $SU(2)$, the 
representations of $SL(2,C)$ are not equivalent to the
complex conjugate representations.  This means that 
we could alternatively replace (\ref{eq:a11}) by  
\begin{equation}
\vert (m,j) \mathbf{P}, \dot{b} \rangle := \sum_{\mu}
\vert (m,j) \mathbf{P}, \mu \rangle_a 
D^j_{\mu \dot{b}}[B_a^{\dagger}(P/M)].
\label{eq:a11b}
\end{equation}
and (\ref{eq:a12}) by 
\begin{equation}
{U(\Lambda, 0)} \vert (m,j) \mathbf{P}, \dot{b} \rangle =
\sum_{{b}'} \vert (m,j) \pmb{\Lambda}{P}, \dot{b}' \rangle
{D^j_{\dot{b}'\dot{b}}[((\Lambda)^{\dagger})^-1)]}.
\label{eq:a12b}
\end{equation}
This 
gives a representation of the scalar product that has the same
form as (\ref{eq:a13}) with the replacement
\begin{equation}
D^j_{bb'}[P^{\mu}\sigma_{\mu}/{M}] \to 
D^{j}_{\dot{b} \dot{b}'}[P^{\mu}
\sigma_2\sigma^*_{\mu}\sigma_2/{M} ] . 
\label{eq:a16}
\end{equation}
While the inner products in all three representations are identical,
the Lorentz covariant and its complex 
conjugate representations are related by
space reflection.  Space reflection changes the kernel
of the Hilbert-space scalar product in the covariant representations.
Space reflection can be represented as an operator on states
by replacing the representations (\ref{eq:a11}) and (\ref{eq:a11b}) 
by a direct sum of both representations. In the direct sum representation the 
wave function becomes a $2 \times (2j+1)$ component spinor
\begin{equation}
\psi ({P} , b) \to 
\left (
\begin{array}{c} 
\xi ({P} , b) \\
\chi ({P} , \dot{b})
\end{array}
\right ) 
\label{eq:a17}
\end{equation}
and the kernel of the inner product becomes
\begin{equation}
d^4P \theta(P^0) \delta (P^2 +
{M^2}) 
\left (
\begin{array}{cc}
D^j_{bb'}[P^{\mu}\sigma_{\mu}/{M}] & 0 \\
0 & D^{j}_{\dot{b} \dot{b}'}[P^{\mu}
\sigma_2\sigma^*_{\mu}\sigma_2/{M} ] 
\end{array}
\right ).
\label{eq:a18}
\end{equation}.  

One desirable feature of the Lorentz covariant representation is that the
basis-dependent features are hidden in the wave functions.  To see
this note that for rotations the upper and lower components have
identical transformations laws
\begin{equation}
U(R, 0) \vert (M,j) \mathbf{P}, {b} \rangle =
\sum_{\dot{b}'} \vert (M,j) R\mathbf{P}, {b}' \rangle
D^j_{{b}'{b}}[R]. 
\label{eq:a19}
\end{equation}
and 
\begin{equation}
U(R, 0) \vert (M,j) \mathbf{P}, \dot{b} \rangle =
\sum_{\dot{b}'} \vert (M,j) R\mathbf{P}, \dot{b}' \rangle
D^j_{\dot{b}'\dot{b}}[R]. 
\label{eq:a20}
\end{equation}
which is the standard rotational transformation law 
that leads to the standard relation  
\begin{equation}
\mathbf{J} = \mathbf{X} \times \mathbf{P} + \mathbf{j}
\label{eq:a21}
\end{equation}
in the covariant representation.

In this representation the relation between the spin, angular momentum,
and orbital angular momentum looks very much like the corresponding
non-relativistic quantities.  The price paid for this simplification
is that the Hilbert space inner product has a non-trivial kernel.
This kernel contains all of the dynamical effects discussed in the
context of Poincar\'e irreducible spins.  The covariant spin is related
to the Poincar\'e irreducible spin of a particle by a boost.  For spin
$1/2$ particles the usual $u$ and $v$ spinors are direct sum
representations of a Lorentz boost.  The choice (helicity, canonical,
light front spin) appears in the representation of these spinors.
The Poincar\'e irreducible labels are the parameters that normally label 
asymptotic states in the $S$-matrix. 

In the end there are many different kinds of spin observables.
In order to measure the spin we need to know how the various 
spin operators couple to the electroweak current operators. 
This will be different for each type of spin observable.

The conclusion is that for relativistic systems the coupling of spins
and orbital angular momenta involves momentum and mass-dependent
Poincar\'e group Clebsch-Gordan coefficients.  In QCD even the parton
masses that appear in these coefficients become variables.  The result
is that the dynamics cannot be ignored in attempts to identify the 
different terms the contribute to the hadronic spin.   

In addition, there are many different spin and orbital angular
momentum observables.  How these different quantities contribute to
the hadronic spin is representation dependent.   As a consequence, 
is important to know how these operators are precisely defined and 
how they are related to experimental observables.

This research was supported by
the US DOE Office of Science, under grant number No. DE-FG02-86ER40286

\end{document}